\documentclass{icrc}

\usepackage{times}
 \usepackage{graphicx} % when using Latex and dvips
\begin{document}

\title{Angular Resolution of Pachmarhi Array of Cerenkov Telescopes}
\author[1]{P.Majumdar}
\author[1]{B.S.Acharya}
\author[1]{P.N.Bhat}
\author[1]{V.R.Chitnis}
\author[1]{M.A.Rahman}
\author[1]{B.B.Singh and P.R.Vishwanath}
\affil[1]{Tata Institute of Fundamental Research,Colaba\\
Mumbai,400005, India}
%\author[2]{C. Authorc}
%\affil[2]{Affililation of Authorc}

\correspondence{pratik@tifr.res.in}

\runninghead{Majumdar et al: Angular Resolution of PACT}
\firstpage{1}
\pubyear{2001}

%\titleheight{11cm} % uncomment and adjust in case your title block
                     % does not fit into the default and minimum 7.5 cm

\maketitle

\begin{abstract}
Pachmarhi Array of \v Cerenkov Telescopes(PACT), \\
consisting of a distributed
array of 25 telescopes
is used to
sample the atmospheric \v Cerenkov Photon showers.
The  shower front is fitted to a plane and the
direction of arrival of primary particle is obtained.
The accuracy in the estimation of the  arrival direction of showers
has been estimated to be $\sim 0^{\circ}.1~$ using `split' array method.
The angular resolution is expected to be even better when a spherical front
is used for direction reconstruction or correction for the curvature of the
front is applied. This is the best angular resolution among all the currently
operating atmospheric \v Cerenkov telescopes in the world.
\end{abstract}

\section{Introduction}
Atmospheric \v Cerenkov Technique (ACT) is a well established and unique method
for the investigation of celestial TeV $\gamma -$ rays. It is mainly based
on the effective detection of \v Cerenkov light emitted by the secondary
particles produced in the extensive air showers initiated by a primary
$\gamma -$ ray[\cite{Cro93}] and reconstructing its direction of arrival in space accurately.
The signal to noise ratio of such an experiment[\cite{Ach93}] is given by
\begin{equation}
S/N \sim  \sqrt{AT/F_{c}\phi} F_{\gamma}
\end{equation}
where A is the physical area of the telescope, T is the time of observations,
$\phi$ is the field of view of the telescope,
 $F_c$ is the background cosmic ray flux and $F_\gamma$ the
flux of
$\gamma-$ rays from the source. In order to achieve high S/N one can either
increase the numerator or decrease the denominator in equation(1). For a given
exposure and resources one can possibly increase S/N by optimising $\phi$, the
telescope aperture, keeping in mind the finite opening angle of the \v Cerenkov cone.
It is possible if the direction of arrival of the shower is
determined accurately, i.e. the error in the estimation of arrival angle has to be
very small or angular resolution has to be high.
The arrival direction of a shower is determined by measuring the relative
arrival time of \v Cerenkov photon front at each of the spatially sparated
telescopes accurately and
reconstructing the shower front. The angular resolution [\cite{Sin87}]
is given by
\begin{equation}
 \delta\theta = (c \delta t)/ (D~cos \theta)
\end{equation}
where $\theta$ is the zenith angle, D the average distance between the telescopes
and $\delta$t the accuracy in timing measurement. The
two factors which contribute to $\delta$$\theta$ are the average
distance, D, between the telescopes and the uncertanity in the measurement of
arrival time of photons. So, if we have a large number of telescopes
separated by large distances, which measure the relative arrival time of photons,
then the shower front could be reconstructed and the direction of the arrival of the
shower could be estimated fairly accurately.
Here we describe
the method of analysis adopted for the estimation of arrival direction of the
 incident primary using
Pachmarhi Array of \v Cerenkov Telescopes (PACT), which is currently in
operation.

\section{Pachmarhi Array of \v Cerenkov Telescopes}
 The experimental set-up of PACT has been explained in detail
elsewhere [\cite{PACT}]. Briefly, it
consists of a 5 x 5 array of atmosphereic \v Cerenkov telescopes deployed
in the form of a rectangular matrix with a separation of 25 m in the N-S
direction and 20 m in the E-W direction.
         Each telescope consists of 7 parabolic mirrors of 0.9 m
diameter with a focal length of 90 cm. Each mirror is viewed by a fast EMI 9807B photomultiplier
tube behind a 3$^\circ$ circular mask. The movement of the telescopes is remotely controlled by a low
cost control
system called Automatic Computerized Telescope Orientation System
(ACTOS) [\cite{ACTOS}].
The alignment of the mirrors is checked with a
bright star (typically of $m_{v}$ $\sim$ 2 to 3) scan.
 Using this method it is ensured that the optic axes of all
the 7 mirrors(labelled A to G) in a telescope are parallel to each other
within an error of
about $0^\circ.2$
The system can orient to the putative source
with an accuracy of $0^\circ.003 \pm 0^\circ.2$. The source tracking is
monitored with an accuracy of $0^\circ.05$ and corrected in real time.
\par The array has been divided into 4 sectors with six telescopes in each
~\footnote{central telescope is presently not included}and the
data are acquired in each sector as well as in central master signal processing
centre separately.
The pulses from 7 PMTs in a telescope are added linearly
to form a telescope trigger pulse called `royal-sum' pulse. Each
`royal-sum' pulse from all the 6
telescopes in a sector are suitably discrminated to yield a count rate of $\sim$ 30-40 kHz.
An event trigger is generated by a coincidence of any 4 of the 6 telescope triggers
in a sector which gives an event trigger rate of $\sim$ 2-5 Hz.
In each sector the timing and density information of \v Cerenkov photons incident on the
6 peripheral mirrors of six telescopes as well as the
timing information on six `royal-sum' telescope pulses  are recorded.
Also, in the
central control room the relative arrival times of all 24 telescope trigger
pulses and sectorwise house-keeping information are recorded.

\subsection{Estimation of Timing Resolution}
The accuracy in timing measurement, $\delta$t,
is estimated as follows. To determine the arrival time of photons
accurately a fast low noise photomultiplier with high gain and minimum
timing jitter is used.
The intrinsic timing jitter of the
signals from the PMTs limit the resolution of timing measurements(0.8 ns).
The event trigger is used as a start pulse to the fast time to digital convertors(TDC).
The individual PMT and `royal-sum' pulses are delayed using ECL
based delay generetors and then fed as TDC stops.
The TDC modules (LeCroy and Philips Scientific make)
were operated with a full scale setting of 500 ns which means a delay of
0.25 and 0.2 ns per count respectively.
Data were collected with all telescopes in the vertical direction. The variance
$\sigma_{ij}$ or
the width of the distribution of difference in relative
arrival times of signals ($\delta t_{ij}$) of
respective TDC channels is an indication of
the limiting accuracy of timing measurement, provided the signals originate
from PMT's located nearby.[\cite{VRC99}]
To minimise the
effects due to fluctuations in the arrival time of
\v Cerenkov photons, which depends upon the  core distance
[\cite{VRC99}] only those combinations corresponding to neighbouring
PMT's are considered. Using this method the limiting
accuracy of timing measurement ($<\sigma_{ij}>/\sqrt{2}$) is estimated to be 1 ns.

%Text in subsection.

%\subsubsection{Subsubsection title}

%Text in subsubsection.

\section{Estimation of Arrival Direction of a Shower}
 The arrival direction of a shower is determined by measuring the relative
arrival time of \v Cerenkov photon front at each telescopes accurately and
reconstructing the shower front. A spherical shape represents the \v Cerenkov
photon shower front fairly accurately, as also demonstrated from Monte Carlo
simulations [\cite{VRC99}]. However, the algorithm and the analysis technique
to determine the shower core in our experiment is under development[\cite{VRC2k1}]. 
In the
absence of the knowledge on the shower core, we assume the
front to be a plane and fit the measured
relative arrival time of \v Cerenkov photons to a plane, normal to which
gives the direction of shower axis. Such an assumption introduces a systematic
error in the
estimation of arrival direction.

\subsubsection{Calculation of Time-offsets}
  The relative arrival time of pulses as measured in the experiment is not
the relative arrival time of \v Cerenkov photons at the PMT, which is needed
for reconstructing the shower front.
A finite but constant delay between pulses from different PMT's (Channels)
arise due to unequal cable lengths, differences in electronic propagation
delays and differences in photomultiplier transit time etc. These are termed as
T0 or Time-offsets. Thus the measured relative arrival times have to be
corrected for this time-offsets to get the relative arrival time of
\v Cerenkov front at the PMT.
The average relative time delays between two PMT's(or telescopes), from a large sample of data,
is entirely due to difference between the two time-offsets. In our
experiment, the average separation between PMT's in a telescope is of the
order of a metre but the separation between the telescopes in a sector is
about 35 m small enough to be ignored. The difference in RMS
fluctuations in arrival time of photons is negligible at this distance
separation for any core distance.[\cite{VRC99}]
If $T0_i$ and $T0_j$ are the time offsets for the PMTs i and j, we can
write an equation of the form
\begin{equation}
 (T0_i - T0_j)  = C_{ij}
\end{equation}
where $C_{ij}$ is the mean delay between a pair of PMT's i and j after
correcting for the time difference due to difference in height
(Z-coordinates) of PMT's if any.

For each pair of PMT's an equation of the form

\begin{equation}
\chi^{2}=\Sigma W_{ij} (T0_i - T0_j -  C_{ij})^{2}
\end{equation}

 can be written, where $W_{ij}$s are the statistical weight factors,
($W_{ij}$=1/${\sigma_{ij}}^{2}$), where $\sigma_{ij}$ is the uncertainity in
determining $C_{ij}$. Using $\chi^{2}$ minimisation one gets an estimate of these
Time-offsets.

\subsection{Reconstruction of arrival direction}
Using the plane front approximation the arrival direction of the shower is
estimated as follows [\cite{Sin87}][\cite{Ach93}].
 If $x_i$,$y_i$,$z_i$ are the coordinates of the
$i^{th}$ PMT, $(l,m,n)$ the direction cosines of the shower axis and
$t_i$ the arrival time of the photons at this PMT then the equation relating
them is given by
\begin{equation}
   lx_i+my_i+nz_i+c(t_i-t_0)=0
\end{equation}
     where $t_0$ is the time at which the shower front passes through the
origin of the coordinate system. Then the arrival direction of the shower
can be estimated by a $\chi^{2}$ minimisation where
\begin{equation}
\chi^{2}=\Sigma w_i ( lx_i +my_i +nz_i +c(t_i - t_0) )^{2}
\end{equation}
where $w_i$ is the statistical weight factor given to the $i^{th}$ timing
mesurement. The values of $(l,m,n)$ and $t_0$ are calculated using the
equations
{$\delta\chi^{2}/\delta l = 0$}, {$\delta\chi^{2}/\delta m = 0$},
{$\delta \chi^{2} / \delta t_0 = 0$} and \(l^{2}+m^{2}+n^{2}=1\).
Fig.1 shows the zenith and azimuthal angle distributions of the
reconstructed arrival directions using the procedure explained above. The reconstruction
of shower front was done using 24 telescopes for the data collected with telescopes in the
vertical position.

\begin{figure}[t]
% \vspace*{2.0mm} % just in case for shifting the figure slightly down
\includegraphics[width=8.3cm]{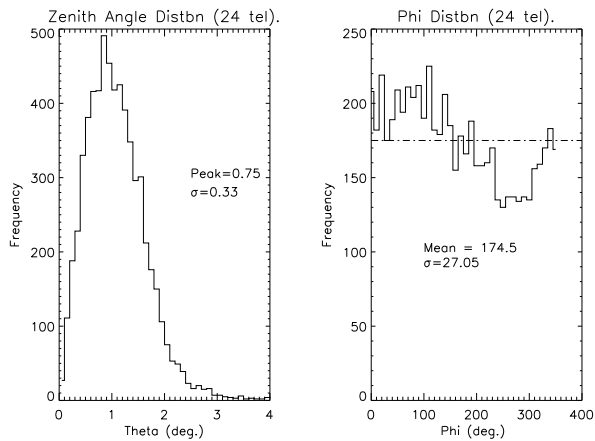} % .eps for Latex,
%                                            % pdfLatex allows .pdf, .jpg, .png and .tif
\caption{ Zenith and Azimuth Angle Distributions for 24 telescopes} 
\end{figure}
\section{Angular Resolution of PACT}
\subsection{ Using Royal Sum TDC Information}
     The angular resolution of PACT has been estimated by using the {\it divided
array} method. In the data recorded in the central control room one has
information on 24 telescopes. The array is divided into two independent parts
of 12 telescopes each,
 say Sectors 1 and 2  and Sectors 3 and 4.
The arrival direction is estimated for each shower from these two independent arrays.
 The distribution of space angle between these two estimates is a measure
of the accuracy with which one can estimate the arrival direction (Figs 2).
Since one has
two independent estimates of the direction the angular resolution will
be given by the peak of the distribution of space angle between the two directions
as \(Peak \over \sqrt{2}\). Similar analysis was performed by dividing the array
using various combinations of telescopes
to study the dependence of angular resolution as a function of max. separation between
telescopes and number of telescopes used in the fit. The results are summarised in
Table \# 1. The term `Odd-Even' refers to 3 telescopes each from Sector 3 and 4 grouped
into one set and the remaining 6 grouped into another.
\begin{table}[hbt] %\centering
\caption{Angular Resolution of PACT using Royal Sums From Control Room Data}
\begin{tabular}{|l|l|l|l|l|l|}
\hline
No.of &Combination& \multicolumn{2}{c|}{Separation}  & Peak of  &Angl\\
tel.&of Detectors &\multicolumn{2}{c|}{bet. Det.(mts)}&Sp.Ang&Resl(deg)\\ \hline
% used  &         &                                           &                   &     \\ \hline
          &                &   Maxm &  Avg     &                  &              \\  \hline
6     & Sec 3 v/s 4 &    53.85  &         31.77      &   0.875  &      0.618          \\ \hline
6     & Odd v/s Even   &   94.33  &          48.81       &    0.637  &      0.45         \\ \hline
12     & Sec 3 and 4 v/s&          &               &            &                  \\
      & Sec 1 and 2 &   94.33  &         44.2       &      0.586  &  0.395         \\ \hline
12     & Sec 1 and 3 v/s &          &               &              &               \\
      & Sec 2 and 4 &  128.07  &         64.56       &      0.47   &  0.3       \\ \hline
\end{tabular}
\end{table}

It is seen from Table \# 1 that as the separation between the detectors increases
there is a definite improvement in the angular resolution  for the same number
 of degrees of freedom.
\subsection{Using Individual Mirror Information}
 The relative arrival times at individual PMT's are available only within
a sector. Data from Sector 3 and 4 and only those events with information in all
6 telescopes are used for the estimation of angular resolution. The angular resolution
is obtained by dividing the data from the sectors into 2 subsets and obtaining the
space angle between the two arrival directions corresponding to two subsets of data.
Information for the 3 PMT's(labelled A, C and E) are grouped into one set while those
for the remaining 3 PMT's of a telescope(B, D and F) are grouped into other set. The
results are summarised in Table \#2. The rows 1 and 4 correspond to the cases in which
two subsets are obtained by demanding all A, all B etc. Rows 2 and 5 correspond to cases
in which the 1st group consists of valid TDC information for any one of 3 A, C, E PMT's
and the second group from any one of 3 B, D, F PMT's. Similarly the rows 3 and 6 correspond
to two sets with at least 2 valid TDC's in each telescope. Finally the row 7 refers to the
case in which the arrival directions are obtained separately from Sector 3 and Sector 4
events and collating event arrival times to pick common events. Column `2' shows the
corresponding no. of
detectors used in the fit for all cases.
Many cases have been studied to understand the improvement
 in angular resolution with increasing degrees of freedom starting from  at
least 1 mirror in a telescope to
greater than 25 in a sector.

\begin{table}[hbt] %\centering
\caption{Angular Resolution of PACT using Individual PMT Information}
\begin{tabular}{|l|l|l|l|l|}
\hline

Sector  & No. of & Combination  & Peak of  & Ang. \\
\#     &  Det. used     & of Detectors &Sp. Angl.  & Resln(deg)        \\ \hline

 3        &    6         &   all A, all B, etc   &         0.46           &   0.325              \\ \hline     3      &     $\geq$ 6         &  at least 1 in &   0.51           &   0.36                \\
          &                  &  each telescope &                   &                  \\ \hline
  3        & $\geq$ 12    &  $\sim$2 in a telescope & 0.43           &  0.3                \\ \hline
  4       &       6       &  all A, all B, etc         &    0.48          &  0.339                \\ \hline
  4       &  $\geq$ 6    &   at least 1 in &                       &                   \\
          &              &     each telescope     &     0.39         &  0.275                 \\ \hline
  4       &  $\geq$ 12   &   $\sim$2 in a telescope & 0.34         &    0.24               \\ \hline
3 and 4   &  $\geq$ 25   &   3,4 Collated by time  &   0.325       &    0.23              \\ \hline

\end{tabular}
\end{table}

  Table \#2 shows that in both the sectors as the number of detectors used in the fit
go up the angular resolution does improve.
On an average 13 to 14 mirrors in a sector have valid TDC information available for
fitting the angle
and hence increase in n, the number of detectors is $\sim$ 2.2. From the angular
resolution computed
as one goes from 6 to 12 mirrors it is seen that the improvement in angular
resolution goes as $\sim$ $n^{0.3}$.
A conservative estimate for the angular resolution($\psi$) of the array is obtained from
Table 2 which is
    $\psi$ = $0.23^\circ$/$2^{-0.75}$ $\times$ $4^{0.3}$ or $0.09^\circ$.

\begin{figure}[t]
% \vspace*{2.0mm} % just in case for shifting the figure slightly down
\includegraphics[width=8.3cm]{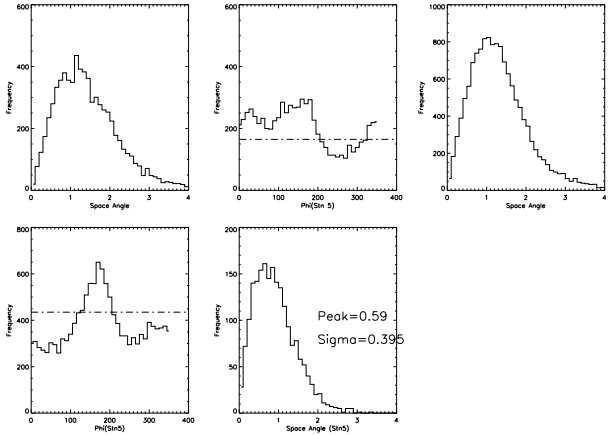} % .eps for Latex,
%                                            % pdfLatex allows .pdf, .jpg, .png and .tif
\caption{ Space Angle Distbn using Sector 5 data. The fit made with Sector
1,2 and Sector 3,4 telescopes.}
\end{figure}

\section{Discussions and Conclusions}
  We have made a detailed analysis on the angular
resolution of PACT using data collected with telescopes in
vertical position. The improvement in angular resolution with longer
baseline between detectors and with increase in the number of degrees of
freedom has been established. The angular resolution of PACT
has been estimated to be $0^\circ.24$ using royal sum TDC information.
The angular resolution from a sector has been
estimated to be $0^\circ.23$ using individual PMT information.
So a conservative estimate of the array is $0.09^\circ$. This is the best angular resolution
achieved so far in the world among all the contemporary atmospheric \v Cerenkov telescopes.
While the angular resolution of the imaging telescopes are limited by the PMT sizes, only
the future imaging telescope arrays (like VERITAS [\cite{veritas}] or HESS) claim a better
angular resolution.
PACT is able to achieve this because of the multiple sampling technique in a distributed
array of ACTs.

%Text of another section.

%New paragraph is preceded by blank line.

%Inline mathematical symbols $a$ or expressions $a=b$ followed by a
%displayed equation
%\begin{equation}
%c=d
%\end{equation}
%followed by more text.

%Vectors should be in bold italics with command $\vec{a}$, matrices
%in bold roman with command $\mathbf{A}$, both in math mode. Units
%should be in roman, either outside the math mode or with
%$\mathrm{cm}$ in math mode.

%Three typs of figure inclusion are demonstrated. Comment out the
%appropriate commands.
% (i) one column figure, will be floated to top of next column
%
\begin{figure}[t]
% \vspace*{2.0mm} % just in case for shifting the figure slightly down
%\includegraphics[width=8.3cm]{detbe3.eps} % .eps for Latex,
%                                            % pdfLatex allows .pdf, .jpg, .png and .tif
% \caption{Figure caption text.}
\end{figure}

% (ii) two column figure, will be floated to top of next page
%
% \begin{figure*}[t]
% \includegraphics[width=17.0cm]{figfile.eps}
% \caption{Figure caption text.}
% \end{figure*}

% (iii) 1 1/2 column figure with caption on the side, will be floated
% to top of next page
% \begin{figure*}[t]
% \figbox*{}{}{\includegraphics*[width=11.0cm]{figfile.eps}}
% \caption{Figure caption text.}
% \end{figure*}

%\section{Balancing}

%The columns of the last page can be balanced either by using the
%command \verb/\balance/ somewhere in the first column of the last page
%or by explicitely put \verb/\vadjust{\newpage}/ at the correct place.
% without the \verb/ / command

\begin{acknowledgements}
It is a pleasure to thank Sarvashri A.I.DSouza, J.Francis, K.S.Gothe, 
B.K.Nagesh,
M.S.Pose, P.N.Purohit, K.K.Rao, S.K.Rao, S.K.Sharma, A.J.Stanislaus,
P.V.Sudershanan, S.S.Upadhyay, B.L.Venkatesha Murthy for their participation 
in various
aspects of the experiment.

\end{acknowledgements}

%\appendix

%\section{Appendix section 1}

%Text in appendix.

\end{document}